# Reproducibility of high-throughput density-functional-theory calculations

LU Chenxi[1], LI Musen[2,3], REIMERS Jeffrey R[1,3]

1. International Centre for Quantum and Molecular Structures and Department of Physics, Shanghai University, Shanghai 200444, China.
2. School of Materials Science and Engineering & Materials Genome Institute, Shanghai University, Shanghai 200444, China
3. University of Technology Sydney, School of Mathematical and Physical Sciences, Ultimo, New South Wales 2007, Australia.

**Abstract:** While standard computational protocols for density functional theory (DFT) have universal applicability, differences exist in code implementations. Specific applications require manual parameter optimization, whereas high-throughput calculations employ predefined workflows. This paper uses the bandgap as a key property to reveal the impact of computational workflow differences on the reproducibility of high-throughput calculation results. The study proposes basic requirements for ensuring reproducibility: using structures optimised using the same procedure as used to calculate properties and ensuring Brillouin zone (*k*-point) integration grid accuracy. This research establishes a foundation for the reproducibility of DFT calculations and reliable application of results, which is of great significance for method development and artificial intelligence model training.

# 高通量密度泛函理论计算的可重复性

鲁晨曦[1]，李木森[2,3]，REIMERS Jeffrey R[1,3]

（1. 上海大学 量子与分子国际中心，上海 200444；
2. 上海大学 材料基因组工程研究院，上海 200444；
3. 悉尼科技大学 数学物理学院，悉尼 2017；）

**摘要**：密度泛函理论(density-functional-theory，DFT)理论的标准计算方案虽具有普遍适用性，但在代码实现中存在差异。特定应用需手动优化参数，而高通量计算采用预设流程。本文以带隙为关键性质，揭示了计算流程差异对高通量计算结果可重复性的影响。研究提出保证可重复性的基本要求：使用密度泛函预测的结构而非实验结构，以及保证布里渊区积分网格精度。本研究为 DFT 计算的可重现性和结果可靠应用奠定了基础，对方法开发和人工智能模型训练具有重要意义。

**关键词**：可重现计算；高通量计算；密度泛函理论；

**关键词**：可重现计算；高通量计算；密度泛函理论；

**中图分类号**：O469    **文献标志码**：A

## 1. INTRODUCTION

Reproducible computational methods are needed in all aspects of computational physics, methods that return the correct answer (to within acceptable limits) for the approach used. In materials science, for semiconductors the orbital bandgap is a quintessential property that underpins most computational objectives, including energy-storage batteries[1], displays and absorbers[2], solar cells[3], sensors[4, 5], and optical-fibre technologies[6], as well as in multiferric devices[7]. The bandgap and other properties are most commonly calculated using density functional theory (DFT)[8, 9], as this method can provide useful answers at affordable computational cost.

Application of DFT requires the choice of a density functional, with a large number of options being available, each with their individual strengths and weaknesses. Predictions for each functional also involve the specification of some significant internal computational parameters, the details of which depend on the DFT computational package used but the principles of which are universal. Reliable calculations give results that are independent of the package used and the internal details, with most computational packages predicting very similar properties when their internal parameters are driven towards their convergence limit. Nevertheless, each packages presents its own recommendations for default calculations intended to deliver useful results in general applications, but these defaults could be far from those needed to obtain convergence universally.

In this work, we compare results previously obtained in three sets of high-throughput calculations, each performed using different computational protocols. One set of results is taken from the orbital-bandgap database developed by Borlido et al.[10] involving 472 semi-conducting 3D materials. Another is taken from the similar work of Kim et al.[11]. In both cases, the computational protocols used were optimised (in different ways) for calculation efficiency. We also a consider an alternative database that we and others recently developed[12] after optimising a new set of computational protocols for calculation reproducibility. We identify the similarities and significant differences of the predictions to ascertain the most significant computational parameters needed to perform calculations that are both reproducible and computationally expedient.

## 2. METHODS

### 2.1 Materials database selection

The materials database developed by Borlido et al.[10] contains 472 semiconducting 3D materials selected from within the Materials-Project (MP) database[13], with considerable overlap between these and the smaller set developed by Kim et al.[11]. Also, the database used for the optimisation of calculation reproducibility[12] contained 340 materials selected from within the Borlido database. This selection excluded materials that changed form on crystal structure optimisation, as well as all noble gasses and some materials that were computational inexpedient for use with the PBE0 and HSE06 functionals. Data from the Kim et al. database were excluded for materials involving the elements Ce, Tl, Pb, and Bi, as for these materials either DFT+U had been used or else spin-orbit coupling (SOC) included.

### 2.2 Review of the general procedures used in the previous reproducibility-optimised calculations

In the critical reproducibility-optimised calculations[12], VASP6.4.3[14, 15] was used to perform all calculations, and some test calculations pertaining to outliers were performed herein using the same procedure. Calculations were performed using the general-gradient-approximation functional PBE[16] as well as the hybrid functionals PBE0[17] and HSE06[18]. In summary, the only functional-specific parameter of note was HFRCUT, which was used to evaluate the long-range component of the exchange operator, as is required only for PBE0. This was set to its default value. Most calculations were performed using the DFT algorithm specified by the command ALGO=ALL, with this replaced by ALGO=NORMAL for the few situations for which the default calculations failed to converge. The energy convergence criterion for the self-consistent field electronic-energy minimisation was set to $10^{-6}$ eV. SOC is not included.

## 3. RESULTS

Metrics are required to understand the difference found between calculations using reproducibility-optimised protocols [12] and the calculations of Borlido et al.[10] and/or Kim et al.[11]. These metrics allow differences to be categorised as being within some bound associated with known calculation shortcomings, and much larger differences beyond expectations that point towards some other significant calculation issue. We consider the parameters identified as being significant to calculation reproducibility, quantification of their likely impact, and then the results comparison and its analysis.

### 3.1 Parameters identified as being key to reproducibility

The focus of the work which optimised DFT protocols for the production of reliable calculations[12] focused on various parameters: (i) the energy cutoff for the plane-wave basis set, $E_{cut}$, which was optimised to 1.3 times the default value, (ii) the pseudopotential used to represent core electrons, with selected the PAW[19] pseudopotentials (PPs) available in VASP that had been optimised for $GW$ calculations, (iii) Monkhorst and Pack[20] $k$-point grids for the integration of the Brillouin zone that had been optimised by numerical interpolation of PBE results obtained using a large $16 \times 16 \times 16$ grid for both the calculation of the valence (HOMO) band edge and the conduction (LUMO) band edge, as well as embodying the "$k$-spacing" method of Wisesa et al.[21] for the evaluation of general properties including general[22], structural optimisation[23], and possible subsequent $GW$[24] calculations, and (iv) the indirect effect of structural optimisation on calculated orbital bandgaps.

### 3.2 Estimated limits of results differences based on known and unknown effects

A summary of the major differences between the reproducibility-optimised protocols and those used by Borlido et al.[10] and/or Kim et al.[11], named #1 - #6, is given in Table 1, along with crude estimates of the maximum errors expected. These estimates are based upon the worst-case errors found in previous high-throughput calculations focusing on the development of reproducible protocols[12], with these results presented only as lower bounds when the methodological differences embody effects beyond those previously considered. Concerning the base error estimates, more details considering whether these occur as random events based on the particular properties of each material, or whether they could be systematically categories as a function of, e.g., atomic composition, are available elsewhere.[12]

Difference #1 concerns the PPs used. The PP is known to effect calculation results[25, 26], with the extensive set used in the reproducibility-optimised calculations having been associated with reproducible results[27, 28]. Compared to the standard protocol of using PBE PP's, an estimate of using 0.16 eV can be made[12] for the maximum-expected orbital-bandgap error in the calculations of Borlido et al. and Kim et al.

Difference #2 concerns the cutoff energy $E_{cut}$. Large values like those used in the reproducibility-optimised calculations[12] lead to good agreement with results obtained using very different computational approaches such as the use of Gaussian basis sets[28]. Table 1 lists a maximum-expected error in the calculations of Borlido et al. and Kim et al. of 0.14 eV [12] associated with this effect.

The choice of structure could have much more profound effects, however. Instead of reproducibly determining the structure predicted by each individual density functional considered, Borlido et al.[10] used experimentally determined structures as listed in the 2019 version of the Materials Project database[29], whereas Kim et al.[11] used structures obtained by optimization using the PBE density functional. It is difficult to estimate the likely consequences of these strategies, and deficiencies in bandgap calculations could exceed the established maximum of 1.2 eV found previously[12] when other approximate approaches were applied. Hence Table 1 lists expected errors of at least 1.2 eV for the calculations of both Bolrido (#3) and Kim (#4).

**Table 1** Possible causes of discrepancies between the calculated results of Borlido et al.[10] and/or Kim et al.[11] and those from reproducibility-optimised calculations[12].a

|    | Calc.     | Limit[a] (eV) | Description                                         |
|----|-----------|---------------|-----------------------------------------------------|
| #1 | All       | 0.16[b]       | Use of lower quality pseudopotentials               |
| #2 | All       | 0.14[b]       | Use of smaller $E_{cut}$                            |
| #3 | Borlido   | > 1.2[b]      | Use of experimentally determined structures         |
| #4 | Kim HSE06 | > 1.2[b]      | Use of PBE optimised structures                     |
| #5 | Borlido   | 0.2[c]        | Similarly or else less optimised $k$-point grid     |
| #6 | Kim HSE06 | ≫ 2.7[b]      | Use of only PBE-generated HOMO and LUMO $k$-points  |

a: Crude estimates for the anticipated largest magnitude of any found difference
b: Largest discrepancy found in previously [12] using pertinent variations for 340 materials.
c: Design criterion for both the reproducibility-optimised calculations[12] and those of Borlido et al.[10] is 0.05 eV, but larger differences are likely, see text.

Concerning the methods used to perform the $k$-space integration, Borlido et al.[10] performed extensive PBE calculations for which their estimated typical error in calculated bandgaps is stated to be 0.05 eV, with the assumption also made that this result applies also to other functionals. This value is also the design criterion for the worst-case error expected from the reproducibility-optimised[12] computational protocols. In the latter case, the design criterion was found to be mostly satisfied, with a few outliers deviating by up to 0.10 eV for PBE and 0.19 eV for PBE0. The PBE result is as expected for a random distribution of errors, but the PBE0 result is excessive and was attributed to unexpected dependencies of the bandgap on the entire *k*-space integration grid owing to the way VASP implements hybrid functionals. The maximum expected differences between the reproducibility-optimised calculations and Borlido's results coming from this effect is therefore set to 0.2 eV in Table 1 (#5).

The HSE06 calculations of Kim et al.[11] were performed using only the vectors determined by PBE to represent the HOMO and LUMO band edges. Hence their HSE06 calculations made no attempt to determine the energy or any other properties of the structures, but would, if orbital energies are fully independent of each other, deliver the anticipated approximate energy for the bandgap. This assumption was found not to hold for hybrid functionals as the way they are implemented in VASP sees information from all *k*-points used to determine the properties of each *k*-point[30]. Errors associated with this effect could be anticipated to exceed significantly the worst-case error found previously[12], 0.2 eV, through consideration of other variations. An additional problem with the method used by Kim et al. is that it assumes that the band edges predicted HSE06 are the same as those predicted by PBE. This assumption is also used in both the reproducibility-optimised protocols and in the calculations performed by Borlido et al., but in these cases if this assumption fails then the methods used provide a backup through the use of extensive *k*-grids that guarantees that a realistic bandgap will be determined. So again it is difficult to estimate a bound for the largest errors possible in the calculations of Kim et al. Here we simply take the largest error found previously [12] associated with *k*-space variations of 2.7 eV as a lower bound for this error, #6 in Table 1.

### 3.3 High-throughput calculation comparison

Table 2 presents a statistical overview of the comparison in calculated orbital bandgaps involving the reproducibility-optimised calculations[12], the calculations of Borlido et al.[10], and the calculations of Kim et al.[11]. This includes the average deviation (AVE), the mean absolute deviation (MAD), the standard deviation (STDEV), as well as the minimum (MIN) and maximum (MAX) deviations between results. The counts of the number of materials compared in each analysis are also listed in the table. Figure 1 shows all of the data used in these analyses, with outliers with absolute bandgap differences exceeding 0.8 eV considered in Table 3. Detailed results are presented in the Supporting Information (SI).

**Table 2** Statistical analyses of differences in bandgaps obtained in reproducibility-optimised calculations[12] with those obtained by Borlido et al.[10] and Kim et al.[11].

| Comparison | | Functional | Count[a] | AVE | MAD | STDEV | MIN | MAX |
| From | to | | | (eV) | (eV) | (eV) | (eV) | (eV) |
| optimised | Borlido | PBE | 331 | 0.00 | 0.15 | 0.21 | -0.94 | 0.59 |
| optimised | Borlido | PBE0 | 255 | -0.20 | 0.19 | 0.26 | -1.55 | 0.32 |
| optimised | Borlido | HSE06 | 252 | -0.10 | 0.16 | 0.23 | -1.36 | 0.37 |
| Borlido | Kim | HSE06 | 171 | -0.07 | 0.22 | 0.46 | -3.43 | 3.15 |
| optimised | Kim | HSE06 | 104 | -0.14 | 0.19 | 0.43 | -3.89 | 0.62 |

a: the number of materials used in the comparison a: Values are listed only if they exceed 0.8 eV in magnitude, with "x" indicating that this comparison was not performed. Hence a blank entry indicates that this method and material was not regarded as being an outlier.

**Table 3** Outliers found in the comparisons with absolute differences greater than 0.8 eV.[a]

| MP-id | composition | Optimised to Borlido | | | Borlido to Kim | Optimised to Kim |
| | | PBE | PBE0 | HSE06 | HSE06 | HSE06 |
| mp-1265 | MgO | | | | -3.43 | -3.89 |
| mp-682 | NaF | -0.81 | -1.23 | -1.09 | | -1.05 |
| mp-7684 | $Tl_3AsSe_3$ | -0.89 | -1.55 | -1.36 | x | x |
| mp-1138 | LiF | -0.94 | -1.52 | -1.24 | x | x |
| mp-569224 | $I_2Sb_4$ | | x | x | -2.49 | |
| mp-1249 | $Mg_2F_4$ | | -0.93 | | | -0.86 |
| mp-856 | $Sn_2O_4$ | | | | -0.85 | -0.92 |
| mp-2815 | $Mo_2S_4$ | | -0.92 | -0.85 | x | x |
| mp-3666 | $Li_2Ta_2O_6$ | | x | -0.88 | x | x |
| mp-555874 | $Li_2As_2S_4$ | | x | x | -0.87 | |
| mp-13682 | $Pd_4S_8$ | | -0.87 | | x | x |
| mp-7262 | $Zn_8As_{16}$ | | x | x | 3.15 | |

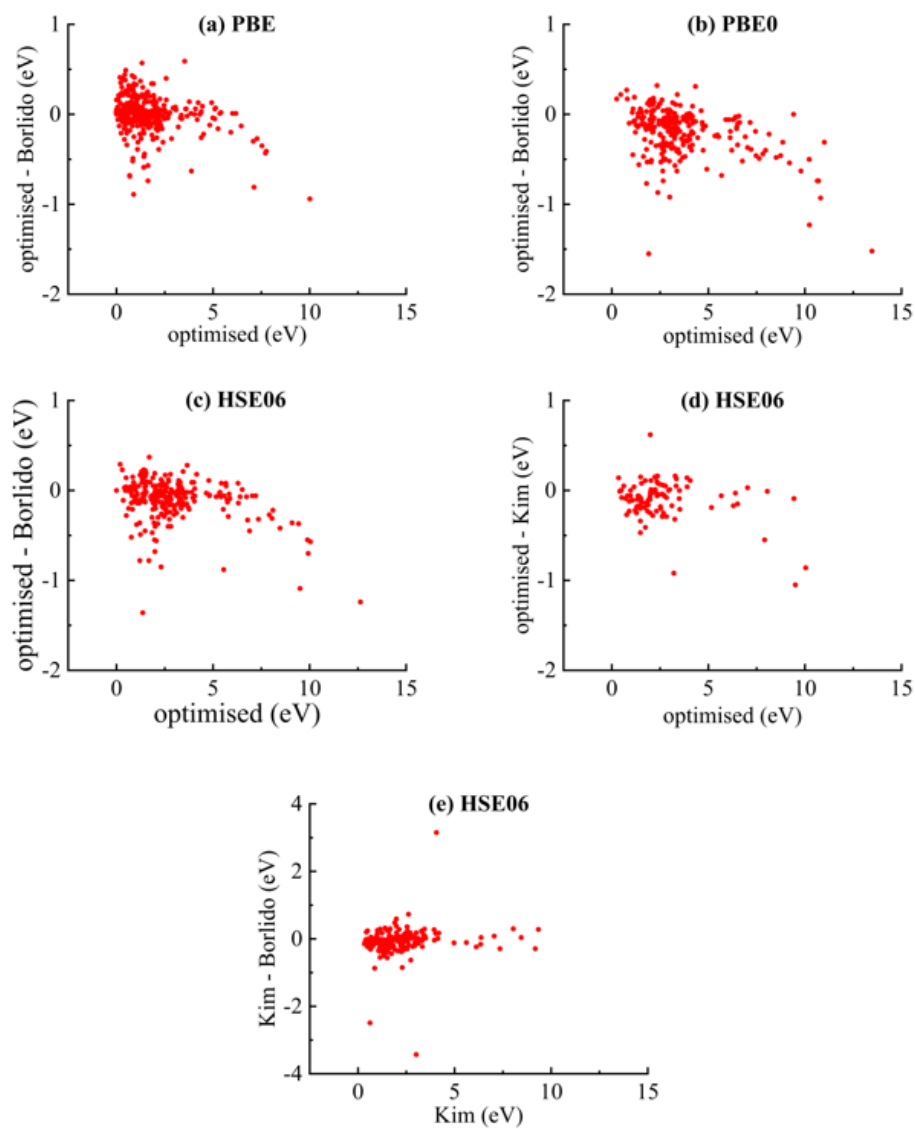

**Fig. 1** Comparison of different DFT bandgap calculations leading to the statistical analysis presented in Table 2. Amongst them, (a) the difference between the reproducibility-optimised PBE calculations[12] and those of Borlido et al.[10]; (b) the difference between the reproducibility-optimised PBE0 calculations[12] and those of Borlido et al.[10]; (c) the difference between the reproducibility-optimised HSE06 calculations[12] and those of Borlido et al.[10]; (d) the difference between the reproducibility-optimised HSE06 calculations[12] and those of Kim et al.[11]; (e) the difference between the HSE06 calculations of Kim et al.[11] and those of those of Borlido et al.[10].

The AVE and MAD differences comparing the reproducibility-optimised[12], Borlido et al.[10] and Kim et al.[11] results for the PBE functional are relatively low, 0.00 eV and 0.15 eV, respectively. This indicates that all calculations are generally consistent. Nevertheless, the SDVEV is significant at 0.21 eV, with the smallest and largest differences being -0.94 eV and 0.59 eV, respectively. This indicates that, for some materials, methodological changes can have profound impact. From the consideration of possible causes of discrepancies between calculations presented in Table 1 the only likely cause of differences of his magnitude is the use of approximate structures by Borlido et al. (#3). Three materials, MgO, NaF, and $Tl_3AsSe_3$ are listed in Table 3 as having large differences for this comparison. Indeed, the way the crystal structures are represented differ between the optimised and Borlido calculations, with significant structural changes evident, see SI.

For the comparison of the reproducibility-optimised calculations[12] with those of Borlido et al.[10] for PBE0 and HSE06, similar effects are revealed in Tables 2 and 3 as found for PBE, but the magnitudes of most effects increase somewhat, with e.g., the error of largest magnitude becoming -1.55 eV. The differences are based to the negative, indicating that Borlido et al.'s bandgaps are smaller. This result is naively expected if unoptimized geometries are used as geometry optimization usually increases bandgaps. This is consistent with the only type of error listed in Table 2 that was anticipated to produce errors of this magnitude, #3.

Large differences are found in Table 2 for HSE06 between the calculations of Kim et al.[11] and either those of Borlido et al.[10] and/or the reproducibility-optimised calculations[12] of up to 4 eV in magnitude. These differences greatly exceed all errors anticipated in Table 1 for all but error #6. Error #6 is associated with the calculation by Kim et al. of just the HSE06 orbitals for the $k$-points for the HOMO and LUMO optimized by PBE. This approximation could be seriously flawed and lead to errors of the identified magnitude. Highlighted outliers suffering this effect in Table 3 are MgO, $Tl_2Sb_4$, and $Zn_8As_{16}$, for which the used structures differ by at most 5 % in axis-vector lengths, effects too small to account for the large bandgap changes.

Of note, the changes in bandgap prediction for the 12 outlier materials listed in Table 3 result in significant worsening of the agreement between calculation and experiment (see SI) for all cases except $Li_2Ta_2O_6$ and $Pd_4S_8$. Agreement with experiment is not universally a good indicator of the reliability of the calculations as DFT-functional inadequacies and ambiguities in experimental data interpretation also influence the results. For $Pd_4S_8$, all functionals overestimate the $c$ cell length by 11%, an effect that is remedied by adding a dispersion correction such as D3[31] to produce useful agreement between the observed and reproducibility-optimised calculated bandgaps. For $Li_2Ta_2O_6$, optimisation of the crystal structure used in the calculations of Borlido et al. results in subtle internal structural distortions that significantly increase the bandgap, leading, in particular, to the highlighted outlier from the PBE0 calculations. Other generic effects that could induce significant inadequacies in reproducible computational approaches include zero-point motion, thermal effects, entropy effects, and the operation of spin-orbit coupling. In summary, reproducible calculations can deliver significant improvements in results when comparing predictions to experiment, owing to intrinsic qualities in the DFT functionals, but can also enhance deviations from experiment, owing to intrinsic weaknesses in functional design.

## 4. CONCLUSIONS

The results from three sets of high-throughput calculations were analysed based on four differences in the computational protocols used: the PP, the energy cutoff energy $E_{cut}$, the details of the geometry of the material, and the Brillouin-zone ($k$-point) integration. Based upon analysis of calculations for up to 340 3D materials, the broad conclusion is that the commonly applied protocols applied in the works of Borlido et al.[10] and Kim et al.[11] deliver useful results, on average, compared to reproducibility-optimised calculations[12]. Nevertheless, for a sizable fraction of the calculations, significant shortcomings were identified, questioning the overall reproducibility of calculations performed using standard computationally efficient protocols.

The most significant shortcomings were found to be associated with the use of approximate structures and the neglect of the full inclusion of the $k$-space grid in calculations involving hybrid functionals. Even though such approaches can lead to significant reduction in the computational time required, the results are inadequate and their application in high-throughput, or indeed any, DFT calculation cannot be justified. To obtain reproducible results, geometrical optimisation of the structure must be performed using the same computational method as used to calculate properties, rather than approximating these using either experimental data or structures optimised using typically computationally cheaper method. Similarly, the use of assumptions concerning the nature of the HOMO and LUMO band edges, and the irrelevance of calculations made at other points within the $k$-space, should not be used in calculations involving hybrid functionals.

Also found to be important is the simultaneous analysis of reproducible calculations and calculations performed at observed geometrical structures. This allows inadequacies in density-functional design to be exposed, as well as providing challenges to experimental data assignment that could considerably enhance the understanding of the properties of materials.

Finally, we note that the use of crude computational approaches that focus on computational expediency at the expense of reliability present significant issues. In particular, they should not be used in the development of other computational approaches, such as in applications of artificial-intelligence to enhance property prediction.

**Acknowledgments**

We thank the Young Scientists Fund of the National Natural Science Foundation of China (Grant No. 12404276), the Special Funds of the National Natural Science Foundation of China (Grant No. 12347164), the China Postdoctoral Science Foundation (Grant Nos. 2024T170541 and GZC20231535), the Australian Research Council Centre of Excellence in Quantum Biotechnology CE230100021, and National Computational Infrastructure (Australia) for provision of computing resources under NCMAS account d63, supported also by the University of Technology Sydney, and ANUMAS account x89.

**References**

[1]     GOLODNITSKY D, YUFIT V, NATHAN M, et al. Advanced materials for the 3D microbattery [J]. Journal of Power Sources, 2006, 153(2): 281-287.

[2]     RONDINELLI J M, KIOUPAKIS E Predicting and designing optical properties of inorganic materials [J]. Annual Review of Materials Research, 2015, 45: 491-518.

[3]     SHARMA S, JAIN K K, SHARMA A Solar cells: in research and applications—a review [J]. Materials Sciences and Applications, 2015, 6(12): 1145-1155.

[4]     DAVIS C M Fiber optic sensors: an overview [J]. Optical engineering, 1985, 24(2): 347-351.

[5]     GRATTAN K T V, SUN T Fiber optic sensor technology: an overview [J]. Sensors and Actuators A: Physical, 2000, 82(1): 40-61.

[6]     SHARMA P, PARDESHI S, ARORA R K, et al. A review of the development in the field of fiber optic communication systems [J]. International Journal of Emerging Technology and Advanced Engineering, 2013, 3(5): 113-119.

[7]     ZHU Y, CHENG Z, WANG X, et al. Synergistic optimization strategies for the development of multienzymatic cascade system-based electrochemical biosensors with enhanced performance [J]. Biosensors and Bioelectronics, 2025, 274: 117222.

[8]     KOHN W, SHAM L J Self-consistent equations including exchange and correlation effects [J]. Physical review, 1965, 140(4A): A1133.

[9]     HOHENBERG P, KOHN W Inhomogeneous electron gas [J]. Physical review, 1964, 136(3B): B864.

[10]    BORLIDO P, AULL T, HURAN A W, et al. Large-scale benchmark of exchange–correlation functionals for the determination of electronic band gaps of solids [J]. Journal of chemical theory and computation, 2019, 15(9): 5069-5079.

[11]    KIM S, LEE M, HONG C, et al. A band-gap database for semiconducting inorganic materials calculated with hybrid functional [J]. Scientific Data, 2020, 7(1): 387.

[12]    LU C, LI M, FORD M J, et al. Reproducible Density Functional Theory Predictions of Bandgaps for Materials [J]. Computational Condensed Matter, 2025, submitted.

[13]    JAIN A, ONG S P, HAUTIER G, et al. Commentary: The Materials Project: A materials genome approach to accelerating materials innovation [J]. APL Materials, 2013, 1(1): 011002.

[14]    KRESSE G, HAFNER J Ab initio molecular dynamics for liquid metals [J]. Phys. Rev. B, 1993, 47: 558-561.

[15]    KRESSE G, HAFNER J Norm-conserving and ultrasoft pseudopotentials for first-row and transition elements [J]. J. Phys. Condens. Matter, 1994, 6: 8245-8257.

[16]    PERDEW J P, BURKE K, ERNZERHOF M Generalized gradient approximation made simple [J]. Phys. Rev. Lett., 1996, 77: 3865-3868.

[17]    ADAMO C, BARONE V Toward reliable density functional methods without adjustable parameters: The PBE0 model [J]. The Journal of chemical physics, 1999, 110(13): 6158-6170.

[18]    KRUKAU A V, VYDROV O A, IZMAYLOV A F, et al. Influence of the exchange screening parameter on the performance of screened hybrid functionals [J]. J. Chem. Phys., 2006, 125(22): 224106.

[19]    KRESSE G, JOUBERT D From ultrasoft pseudopotentials to the projector augmented-wave method [J]. Phys. Rev. B, 1999, 59: 1758.

[20]    MONKHORST H J, PACK J D Special points for Brillouin-zone integrations [J]. Physical review B, 1976, 13(12): 5188.

[21]    WISESA P, MCGILL K A, MUELLER T Efficient generation of generalized Monkhorst-Pack grids through the use of informatics [J]. Physical Review B, 2016, 93(15): 155109.

[22]    GUO Q, WANG G, BATISTA E R, et al. Two-Dimensional Nanomaterials as Anticorrosion Surface Coatings for Uranium Metal: Physical Insights from First-Principles Theory [J]. ACS Applied Nano Materials, 2021, 4(5): 5038-5046.

[23]    CHEN Z, LI H, ZHANG C, et al. Crystal Structure Prediction Using Generative Adversarial Network with Data-Driven Latent Space Fusion Strategy [J]. Journal of Chemical Theory and Computation, 2024, 20(21): 9627-9641.

[24]    LV Y-Y, ZHOU Y, XU L, et al. Non-hydrostatic pressure-dependent structural and transport properties of BiCuSeO and BiCuSO single crystals [J]. Journal of Physics: Condensed Matter, 2021, 33(10): 105702.

[25]    LEJAEGHERE K, VAN SPEYBROECK V, VAN OOST G, et al. Error estimates for solid-state density-functional theory predictions: an overview by means of the ground-state elemental crystals [J]. Critical reviews in solid state and materials sciences, 2014, 39(1): 1-24.

[26]    JOLLET F, TORRENT M, HOLZWARTH N Generation of Projector Augmented-Wave atomic data: A 71 element validated table in the XML format [J]. Computer Physics Communications, 2014, 185(4): 1246-1254.

[27]    ARYASETIAWAN F, GUNNARSSON O The GW method [J]. Reports on Progress in Physics, 1998, 61(3): 237.

[28]    LI M, REIMERS J R, FORD M J, et al. Accurate prediction of the properties of materials using the CAM-B3LYP density functional [J]. J. Comput. Chem., 2021, 42(21): 1486-1497.

[29]    JAIN A, ONG S, HAUTIER G, et al., The Materials Project: a materials genome approach to accelerating materials innovation, APL Mater. 1 (2013) 011002.

[30]    DUCHEMIN I, GYGI F A scalable and accurate algorithm for the computation of Hartree–Fock exchange [J]. Computer Physics Communications, 2010, 181(5): 855-860.

[31]    GRIMME S, ANTONY J, EHRLICH S, et al. A consistent and accurate ab initio parametrization of density functional dispersion correction (DFT-D) for the 94 elements H-Pu [J]. J. Chem. Phys., 2010, 132: 154104.